\documentstyle[11pt]{article}
\begin{document}
\vspace{20mm}

\title{Solitons on Noncommutative Torus as \\
       Elliptic Algebras and Elliptic Models}
\author{Bo-Yu Hou$$ \thanks{Email:byhou@phy.nwu.edu.cn}, \hspace{5mm}
       Dan-Tao Peng$$ \thanks{Email:dtpeng@phy.nwu.edu.cn}, \hspace{5mm}
       Kang-Jie Shi$$ \thanks{Email:kjshi@phy.nwu.edu.cn}, \hspace{5mm}
       Rui-Hong Yue$$ \thanks{Email:yue@phy.nwu.edu.cn}\\[3mm]
       Institute of Modern Physics, Northwest University\\
       Xi'an, 710069, China}
\date{}
\maketitle

\begin{abstract}
For the noncommutative torus ${\cal T}$, in case of the N.C. parameter
$\theta = \frac{Z}{n}$ and the area of ${\cal T}$ is an integer, we
construct the basis of Hilbert space ${\cal H}_n$ in terms of $\theta$
functions of the positions $z_i$ of $n$ solitons. The loop wrapping around
the torus generates the algebra ${\cal A}_n$. We show that ${\cal A}_n$ is
isomorphic to the $Z_n \times Z_n$ Heisenberg group on $\theta$ functions.
We find the explicit form for the local operators, which is the generators
$g$ of an elliptic $su(n)$, and transforms covariantly by the global
gauge transformation of the Wilson loop in ${\cal A}_n$. By acting on
${\cal H}_n$ we establish the isomorphism of ${\cal A}_n$ and $g$. Then it
is easy to give the projection operators corresponding to the solitons and
the ABS construction for generating solitons. We embed this $g$ into the
$L$-matrix of the elliptic Gaudin and C.M. models to give the dynamics.
For $\theta$ generic case, we introduce the crossing parameter $\eta$
related with $\theta$ and the modulus of ${\cal T}$. The dynamics of
solitons is determined by the transfer matrix $T$ of the elliptic quantum
group ${\cal A}_{\tau, \eta}$, equivalently by the elliptic Ruijsenaars
operators $M$. The eigenfunctions of $T$ found by Bethe ansatz appears to
be twisted by $\eta$.

\vspace{.5cm}

\end{abstract}

\setcounter{equation}{0}

\section{Introduction}

\indent

The development of soliton theory on noncommutative geometry is rather
impressive (details of reference e.g. see review \cite{H, DN}). The paper
\cite{GMS} gives the solitons expressed by the projection operators.
Witten \cite{W} gives the partial isometry operators. The paper \cite{HKL}
uses the partial isometry operators to generate the solitons on
noncommutative plane.

Recently, the solitons on the N.C. torus has attracted a lot of interest
\cite{BKMT, SS, MM, B, GHS, TKMS, KMT, Bars}. On the compactified torus
\cite{CDS}, the duality, the Morita equivalence and the orbifolding have
been studied \cite{SW}.

The equivalence class of projection operator on torus is given by
\cite{R} in terms of $U_1$ and $U_2$. The noncommutative algebra ${\cal
A}$ generated by $U_i$ ($U_1 U_2 = U_2 U_1 e^{i\theta}$) wrapping the
torus has been given in \cite{CDS, SW, KS} in terms of the matrix
difference operator. The matrix acts on a $U(n)$ bundle $V_n$ with trivial
connection while the difference acts on a $U(1)$ bundle ${\cal L}$. The
Hilbert space ${\cal H} = V_n \otimes {\cal L}$ acted  by ${\cal A}$ is
given by the vector functions of {\it real} variables. The local covariant
derivative operator $\bigtriangledown_i$ acted on ${\cal L}$ is given
also. Their commutator gives the constant curvature related to $\theta$.
Obviously the periodicity of wave functions on torus can not be given
explicitly by real variable functions. The Schwarz space of smooth
functions $S({\bf R})$ of rapid decrease real functions provides a
bimodule between ${\cal A}_\theta$ and ${\cal A}_{\frac{1}{\theta}}$ under
Morita equivalence \cite{MM, KMT}. Starting from the Gaussian function
$e^{i\pi y^2/\theta}$ as the Schwarz function, Boca \cite{B} constructed
the projection operators in terms of $\theta$ functions. This projection
operator satisfies the BPS like selfduality condition \cite{BKMT, DKL},
thus gives constant curvature. But Boca obtained the explicit expressions
only for the case with modulus equals $\frac{1}{Z}$, moreover, it is in
terms of the products of two $\theta$ functions depends seperately on
$U_1$ and $U_2$, so the symplectic structure of noncommutative torus is
unclear. Gopakumar et al \cite{GHS} starting from the same Gaussian
function but orbifolding the double periodic multisoliton solution on the
N.C. ${\bf R}^2$ into a single soliton solution on the torus. They
employed the $kq$ representation \cite{BCZ, Zak} on a dual lattices for
conjugate variables $k$ and $q$, which provides a basis of simultaneous
eigenstates of commutative $U_1$ and $U_2$ with $\theta = 2 \pi A$, where
$A$ is an integer. Thus they succeeded in constructing the soliton on the
so-called integral torus with double periodic wave functions. The
noncommutative symplectic complex structure on torus appears explicitly
and the corresponding Weyl-Moyal transformation is realized as double
series of $U_1^{\frac{1}{A}}$ and $U_2^{\frac{1}{A}}$. But since the loop
$U_1$ and $U_2$ become commutative in this degenerated $\theta = 2 \pi A
(e^{i \theta} = 1)$ case, they obtained just a unique projection operator,
there is neither the covariant derivative corresponding to the local
translation nor the ABS like partial isometry operator. Moreover, it is
not obvious how to generalize into the case of generic $\theta$.

This paper is just atempted at first to give explicitly more projection
operators, the local derivative operators $E_\alpha$ and the soliton
generating ABS operators by orbifolding such integral torus ${\cal T}$
into a ${\cal T}_n$; secondly to realize the noncommutative algebra ${\cal
A}_\theta$ in case of generic $\theta$ explicitly on a Hilbert space of
meromorphic functions. We try to relate various algebra on noncommutative
torus, the $Z_n \times Z_n$ Heisenberg algebra with the elliptic
$su_n({\cal T})$ algebra and the elliptic Sklyanin algebra with the
elliptic quantum group, then furthermore consider the dynamics of the
noncommutative solitons by relate it to the well known integrable models
such as the elliptic Gaudin models with the ellipitc Calogera-Moser models
and the elliptic IRF(interaction aroud face) models with the elliptic
Ruijsenaars-Schneider models.

In the next section, after review the construction by Gopakumar et. al
\cite{GHS} for the soliton on the integral torus ${\cal T}$, we show by $n
\times n$ times orbifolding this ${\cal T}$ into ${\cal T}_n$, that the
one dimensional trivial Heisenberg group i.e. the double periodicity under
wrapping by $U_i$ on ${\cal T}$ in paper \cite{GHS} will be refined into a
$Z_n \times Z_n$ $G_{\cal H}(n)$ generated by loops $W_i (\equiv
U_i^\frac{1}{n})$ around ${\cal T}_n$.

In section 3, we use the $\theta$ functions of the central "positions"
$z_i$ of multisolitons to give an explicit expression for the basis of the
Hilbert space ${\cal H}_n$. The ${\cal A}_n$ algebra turns to be the $Z_n
\times Z_n$ Heisenberg group $G_{\cal H}(n)$ acting on $\theta$ functions
\cite{Tata}. The local infinitesimal translation operator $E_\alpha$ is
realized as the derivative operator generating the $su_n({\cal T})$
algebra on torus. The $E_\alpha$ behaves covariantly under gauge
transformation by $G_{\cal H}(n)$. We establish the isomorphism of the
algebra ${\cal A}_n$ and the $su_n({\cal T})$ by applying them on ${\cal
H}_n$, they both become identical to the $G_{\cal H}(n)$. Then it is easy
to construct the ABS and the projection operators corresponding to
multiple solitons.

Later, in section 4, we embed this $su_n({\cal T})$ derivative operator as
the "quantum" operators in the representation of the transfer matrix (Lax
operator) of the elliptic Gaudin model i.e. $L(u)$ is obtaned from the
classical Yang-Baxter $r$-matrix $r(u) = \sum_\alpha w_{\alpha}(u)
I_\alpha \otimes I_\alpha$ acted on $V_n \otimes V_n$ space by replacing
one $V_n$ into a "quantum" space, i.e. replacing the Heisenberg matrix
$I_\alpha$ on it by the $su_n({\cal T})$ operator $E_\alpha$. The
$r$-matrix is nondynamic. It is depends on the spectral (evaluation)
parameter (u) through the meromorphic sections $w_\alpha(u)$. The double
Heisenberg properties if $w_\alpha(u)$ and hence of $r(u)$ ensures $L(u)$
to be a section on a twisted $su(n)$ bundle. Then we gauge transform the
Gaudin Lax into that of the elliptic Calogero-Moser models. This Lax is
equivalent to the elliptic Dunkle operators representing Weyl reflections.
The trace of the quadratic of $L$ gives the Hamiltonian of the C. M. which
is assumed to give the dynamics of the N.C. solitons on the brane. As in
section 3 we have shown that the isomorphisim of N.C. $z_i$ and
$\bar{z}_i$ to the $z_i$ and local translation $\partial_i$ on ${\cal H}$
in N.C. ${\cal R}^2$ case, under orbifold ${\cal R}^2 / Z \otimes Z$
becomes the isomorphism  of the $Z_n \times Z_n$ Heisenberg $W^\alpha$ in
${\cal A}_n$ to the $E_\alpha$ of $su_n({\cal T})$. So the correspondence
of noncommutative $\partial_i (\sim \bar{z}_i)$ and $z_i$ in $E_\alpha$ to
dynamical $p_i$ and $q_i$ in $L(u)$, natrually endow the N.C. torus with
symplectic structure. Thus naturally the solitons on branes satisfy the
dynamics of Calogero-Moser systems.

Then we turn to the case of generic $\theta$. Here, at first in section 5,
we find that, the difference representations of the Sklyanin algebra,
where the difference operation is to shift the solitons position $z_i$ by
$\eta$, will give a constant Heisenberg matrix on the functions in $V_n$
introduced in section 3. The chracteristic determinant of this operator
gives the Ruijsenaars operators. Since the Ruijsenaars-Schneider model in
the limit $\eta \rightarrow 0$ approaches the  C.M. model, it is
reasonable to assume that the dynamics of the soliton on $\theta$ generic
torus is given by using these Ruijsenaars operators as the Hamiltonian.

But this Sklyanin algebra on the vector space $V_n$ gives only the
constant matrix part of the trivial $su(n)$ bundle. It is not sufficient
to show the whole $su(n) \otimes u(1)$ operations as \cite{CDS, SW}. So in
section 6 we realize the full algebra as the elliptic quantum group
$E_{\tau, \eta}$ constructed by Felder. It is well known that the
$E_{\tau, \eta}$ is equivalent to a ${\cal S}_{\tau, \eta}$, their
evaluation representations is related to the elliptic IRF models. At first
we give a factorized form of this representation, which may be gauge
transfromed into the Sklyanin algebra and further we relate its transfer
matrix to the Ruijsenaar operators $M$. These representations of ${\cal
S}_{\tau, \eta}$ are the to be found matrix difference operators which
realize the corresponding operators given by \cite{CDS, SW}, but instead
of acting on real vector functions, now it acts on meromorphic vector
sections. Then as Felder and Varchenko \cite{FV2}, using the matrix
difference operator $M$ to represent the transfer matrix, we apply the
Bethe ansatz to find the common eigenfunctions which actually twisted by
$\eta$ for N.C. $U_i$ and realizes the Hilbert space \cite{CDS, SW} in the
form of the vector meromorphic sections (sheafs).

In the last section, we shortly describe the subjects which will be
investigated later.

\section{Solitons on the "integral torus" \cite{GHS} and its further
orbifolding}

\indent

In this section we at first shortly review the result of paper \cite{GHS}
for solitons on nocommutative space ${\bf R}^2$: $[\hat{x}_1, \hat{x}_2]
= i\Theta$. This ${\bf R}^2$ has been orbifolded to a torus ${\cal T}$
with periodicities $L$ and $\tau L$.

The generators of the fundamental group of ${\cal T}$ are:
\begin{equation}
U_1 = e^{- i \hat{y}^2 l}, \hspace{10mm} U_2 = e^{i l (\tau_2 \hat{y}^1 -
\tau_1 \hat{y}^2)},
\end{equation}
where $l$ is the normalized length $l = L\sqrt{\Theta}$,  $\hat{y}^i =
\frac{1}{\sqrt{\Theta}}\hat{x}^i$, $(i = 1, 2)$ and
\begin{equation}
U_1 U_2 = U_2 U_1 e^{- i\tau_2 l^2}
\end{equation}
As in \cite{GHS}, let us consider the case $\frac{\tau_2 l^2}{2\pi} \in
{\bf N}$ (or ${\bf Z}_+$) i.e. the normalized area $A = \frac{\tau_2
l^2}{2 \pi} = B$ the flux passing through the torus is an integer. The
projection operators constructed by \cite{GHS} has its image spanned by
the lattice of coherent states
\begin{equation}
U_1^{j_1} U_2^{j_2}|0\rangle, (j_1, j_2 \in {\bf Z}^2)
\end{equation}
here $|0\rangle$ is the Fock Bargmann vacuum vector: $a|0\rangle = 0$,
where $a = \frac{1}{\sqrt{2}}(\hat{y}^1 + i\hat{y}^2)$, $a^\dagger =
\frac{1}{\sqrt{2}}(\hat{y}^1 - i\hat{y}^2)$, $[a, a^\dagger] = 1$. They
found a particular linear combination:
\begin{equation}
\label{wavefunction}
|\psi\rangle = \sum_{j_1, j_2}c_{j_1, j_2}U_1^{j_1}U_2^{j_2}|0\rangle
\end{equation}
that satisfies
\begin{equation}
\label{orthonormal}
\langle\psi|U_1^{j_1}U_2^{j_2}|\psi\rangle = \delta_{j_1 0}\delta_{j_2 0}.
\end{equation}
Then the projection operator becomes
\begin{eqnarray}
P & = & \sum_{j_1, j_2} U_1^{j_1} U_2^{j_2} |\psi\rangle
\langle\psi|U_2^{- j_2} U_1^{- j_1} \nonumber\\
P^2 = P.
\end{eqnarray}
Since (\ref{orthonormal}), we have $U_i P U_i^{-1} = P$, $(i = 1, 2)$,
i.e. $P$ actually acts on ${\cal T}=\frac{{\cal R}^2}{{\bf Z}\times {\bf
Z}}$ with periods $U_i$.

To find the solution satisfying the orthonormality condition
(\ref{orthonormal}), the wave function of (\ref{wavefunction}) has been
found \cite{GHS} in terms of the $kq$ representation \cite{BCZ, Zak}:
\begin{equation}
\label{kq}
|kq\rangle =
\sqrt{\frac{l}{2\pi}}e^{-i\tau_1(\hat{y}^2)^2/{2\tau_2}} \sum_j e^{ijlk}
|q + jl\rangle.
\end{equation}
where $|q\rangle$ is an eigenstate of $\hat{y}^1$: $\hat{y}^1 |q\rangle =
q |q\rangle$, such that
\begin{equation}
e^{i l \tau_2 \hat{y}^1}|q\rangle = e^{ilq\tau_2}|q\rangle, \hspace{5mm}
e^{- i l \hat{y}^2}|q\rangle = |q + l\rangle.
\end{equation}
Obviously, in case $A$ being an integer, $|kq\rangle$ is the common
eigenstate of $U_1$ and $U_2$:
\begin{equation}
U_1|kq\rangle = e^{-ilk}|kq\rangle, \hspace{1cm} U_2|kq\rangle =
e^{il\tau_2 q}|kq\rangle.
\end{equation}
Remarks: From the completeness of the coherent states \cite{Perelomor}, we
know that $|kq\rangle$ is defined in the double periods: $0\leq k\leq
\frac{2\pi}{l}, 0\leq q\leq l$, and constitute an orthonormal and complete
basis for the Hilbert space ${\cal H}_A = {\cal H}_{{\bf R}^2}/A$, if $A
\geq 2$. In case of $A = 1$, if we choose just one coherent state for any
unit cell, then all the state excluding the vacuum state maybe taken as a
complete and minimal system of state. Bars \cite{Bars} recently has
discussed the solutions with size smaller than this extremal one.

By using $|q\rangle = \exp(\frac{(a^\dagger)^2}{2} + a^\dagger q +
\frac{q^2}{2})|0\rangle$, we find the wave function
\begin{equation}
\label{kq-wave}
\langle kq | 0 \rangle = \sqrt{\frac{l}{2 \pi}} e^{\frac{\tau}{2 i
\tau_2}q^2}\theta(\frac{(k + i q)A}{\tau l}, \frac{A}{\tau}),
\end{equation}
where $\tau = \tau_1 + i \tau_2$. Then by modular transformation of
(\ref{kq-wave}) or directly from (\ref{kq}) by using the Poisson
resummation formula we obtain as in \cite{GHS}:
\begin{equation}
\label{wave-func}
C_0(k, q) \equiv \langle kq | 0 \rangle =
\frac{1}{\pi^{\frac{1}{4}}\sqrt{l}}\exp(-\frac{\tau}{2i\tau_2}k^2 + ikq)
{\cal \theta}_{00}(\frac{q + k\tau/{\tau_2}}{l}, \frac{\tau}{A}).
\end{equation}
The orthonormality condition (\ref{orthonormal}) becomes
\begin{eqnarray}
\delta_{j_1, 0}\delta_{j_2, 0} & = & \int_0^{\frac{2 \pi}{l}} dk \int_0^l
dq e^{- i j_1 l k + i j_2 l \tau_2 q}|\Psi(k, q)|^2\nonumber\\
& = & \int_0^{\frac{2 \pi}{l}} dk \int_0^{\frac{l}{A}} dq e^{- i j_1 l k +
i j_2 l \tau_2 q}|\tilde{c}(k, q)|^2 \sum_{n = 0}^{A - 1}\left | C_0(k, q
+ n \frac{l}{A})\right |^2.
\end{eqnarray}
Thus, paper \cite{GHS} find that
\begin{equation}
\label{Psi}
\langle k q | \psi \rangle \equiv \Psi(k, q) = \tilde{c}(k, q) C_0(k, q) =
\frac{C_0(k, q)}{\sqrt{2 \pi / A \sum_{n = 0}^{A - 1}|C_0(k, q + n
l/A)|^2}}.
\end{equation}

Now we turn to the further orbifolding. Let $W_1 = U_1^{\frac{1}{n}}$,
then
\begin{equation}
C_1(k, q) \equiv \langle kq | W_1 | 0 \rangle = \langle k, q + \frac{l}{n}
| 0 \rangle = \frac{1}{\pi^{\frac{1}{4}} \sqrt{l}} \exp(- \frac{\tau
k^2}{2 i \tau_2} + i k (q + \frac{l}{n})) \theta_{0, \frac{1}{n}}
(\frac{q + \frac{k \tau}{\tau_2}}{l}, \frac{\tau}{A}),
\end{equation}
here we have chosen $A$ and $n$ relatively prime.

Similarly
\begin{eqnarray}
C_\beta(k, q) \equiv \langle kq | W_1^\beta | 0 \rangle & = &
\frac{1}{\pi^{\frac{1}{4}}\sqrt{l}} \exp(- \frac{\tau k^2}{2 i \tau_2} + i
k (q + \frac{\beta}{n} l)) \theta_{0, 0}(\frac{q +
\frac{k\tau_1}{\tau_2}}{l} + \frac{\beta}{n}, \frac{\tau}{A})\nonumber\\
& \equiv & \frac{1}{\pi^{\frac{1}{4}}\sqrt{l}} \exp(- \frac{\tau k^2}{2 i
\tau_2} + i k (q + \frac{\beta}{n} l)) \theta_{0, \frac{\beta}{n}}(z,
\frac{\tau}{A}),
\end{eqnarray}
here $z \equiv y_1 + i y_2$, the map $k, q$ to $z$ is given \cite{GHS} by
the Weyl Moyal transformation. The effect of $W_1$ is to shift $z$ to $z +
\frac{1}{n}$, or equivalently shift the characteristic $\beta$ of
$\theta_{\alpha, \beta}$ to $\beta + \frac{1}{n}$.

Constructing
\begin{equation}
\label{Psi_beta}
\langle k q |\Psi_\beta\rangle = \frac{C_\beta(k, q)}{\sqrt{\frac{2\pi}{A}
\sum_{n=0}^{A-1}|C_\beta(k, q + n \frac{l}{A})|^2}},
\end{equation}
then $|\Psi_{\beta + n}\rangle = e^{-i l k}|\Psi_\beta\rangle$ and
$|\Psi_\alpha \rangle$ ($\alpha = 0, 1, \cdots, n-1$) are linearly
independent.

Let
\begin{equation}
\label{P_beta}
P_\beta = \sum_{j_1, j_2} U_1^{j_1} U_2^{j_2} | \Psi_\beta \rangle \langle
\Psi_\beta | U_2^{- j_2} U_1^{- j_1},
\end{equation}
then
\begin{equation}
P_{\beta}^2 = P_{\beta}
\end{equation}
\begin{equation}
W_1^{\beta_1} P_{\beta_2} W_1^{- \beta_1} = P_{\beta_1 + \beta_2}.
\end{equation}

As in paper \cite{SW}, We may also orbifold the torus along the
$\overrightarrow{\tau}$ direction $n$ times. Let $W_2 =
U_2^{\frac{1}{n}}$, then
\begin{eqnarray}
C_{\alpha}^\prime(k, q) & \equiv & \langle kq | W_2^{\alpha}| 0 \rangle =
e^{i \frac{\alpha}{n} l \tau_2 q} \langle k + \frac{\tau_2 l}{n}\alpha,
q | 0 \rangle \nonumber\\
& = & \frac{1}{\pi^{\frac{1}{4}}\sqrt{l}}\exp(- \frac{\tau k^2}{2 i
\tau_2} + i (k + \frac{\alpha}{n} l \tau_2) q) \theta_{\frac{\alpha}{n} A,
0}(\frac{q + \frac{k \tau}{\tau_2}}{l}, \frac{\tau}{A}).
\end{eqnarray}
Obviously, $W_2$ shifts $z$ to $z + \frac{\tau}{n}$, i.e. shift the
$\alpha$ of $\theta_{\alpha, \beta}$.

Subsequently, we may construct $|\psi_{\alpha}^\prime \rangle$ and
$P_{\alpha}^\prime$ as (\ref{Psi_beta}) and (\ref{P_beta}). But the set
$P_{\alpha}^\prime$, $(\alpha = 1, 2, \cdots, n)$ are not independent from
the set $P_\beta$, $(\beta = 1, 2, \cdots, n)$, and in either basis the
$W_1$ or(and) $W_2$ matrix are not constant. Actually, the target space
${\cal H}_{\cal T}$ of $P$ on total torus ${\cal T}$ has been subdivided
into ${\cal H}_n$ on torus ${\cal T}_n$ as described by \cite{SW}. One may
find a basis of ${\cal H}_n$ such that $W_1$ and $W_2$ become the $n
\times n$ irreducible matrix representation of the Heisenberg group
\begin{equation}
W_1 W_2 = W_2 W_1 \omega, \hspace{1cm} \omega^n = 1,
\hspace{1cm} W_1^n = W_2^n = 1.
\end{equation}

In next section we will construct explicitly this basis in terms of
$\theta_{\alpha \beta}$ functions. But before that, let us compare to the
case of N.C. plane, that the generic $n$ solitons solution is $\prod_{i =
1}^n(a^\dagger - z_i)|0\rangle$, with $n$ soliton centers $z_i$.
Similarly, we will introduce the $z_i$ for the location of the centers of
$n$-solitons solutions on the torus ${\cal T}$, such that the moduli space
of the $n$-soliton solution is ${\cal T}^{\otimes n}/S_n$.

\section{Noncommutative algebra ${\cal A}_n$, Hilbert space ${\cal H}_n$,
Heisenberg group $Z_n \times Z_n$ and $su(n)$ algebra, in case $[U_1, U_2]
= 0$}

\subsection{The Hilbert space}

\indent

As in \cite{SW}, usually the Hilbert space ${\cal H}_{\cal T}$ can be
written as the direct product of a $su(n)$ trivial bundle $V_n$ and an
$U(1)$ line bundle ${\cal L}$: ${\cal H}_{\cal T} = V_n \otimes {\cal
L}$. On the $V_n$ acts the $W_1$ and $W_2$ matrices, on the ${\cal L}$
acts the covariant defference operators $\hat{\bf V}_i$ (the notation
of \cite{CDS}), and $W_i \otimes \hat{\bf V}_i = U_i$. Thus, the
operators $U_i$ are matrix difference operators acting on vector
functions $v_a, (a = 1, 2, \cdots, n)$. But now it happens that in case
of commutative $U_1$ and $U_2$, this $V_n$ is identical to the whole
${\cal H}_n$ on subtorus ${\cal T}_n$ in the following way.

The basis vectors of the {\it Hilbert space ${\cal H}_n$} are
\begin{eqnarray}
\label{basis}
v_a & = & \sum_{b=1}^n F_{-a, b}, (a= 1, 2, \cdots, n), \nonumber\\
F_{\alpha} & \equiv & F_{\alpha_1, \alpha_2} = e^{i\pi n \alpha_2}
\prod_{j=1}^n \sigma_{\alpha_1, \alpha_2}(z_j - \frac{1}{n}\sum_{k=1}^n
z_k),
\end{eqnarray}
here $\alpha \equiv (\alpha_1, \alpha_2) \in Z_n \times Z_n $, and
$$
\sigma_{\alpha}(z) = \theta \left [
\begin{array}{c}
\frac{1}{2} + \frac{\alpha_1}{n}\\
\frac{1}{2} + \frac{\alpha_2}{n}\\
\end{array}\right ](z, \tau).
$$
(In the following, we use the modulus $\tau$ to represent the
$\frac{\tau}{A}$ in section 2).

Remark: The $\Psi$ (\ref{Psi}) has been normalized, but it is not entire,
i.e. it has poles and is nonanalytic. Our basis $v_a$ is entire, they
span the space of weight $n$ quasiperiodic functions \cite{Tata}, i.e.
the space of $n$ soliton sections of some $su(n)$ bundle.

Now we turn to show that the $W_1^{\alpha_1}, W_2^{\alpha_2}$ with
$(\alpha_1, \alpha_2) \in Z_n \times Z_n$ acting on this ${\cal H}_n$
generates the algebra ${\cal A}_n$ on ${\cal T}_n$.

\subsection{Noncommunitative algebra ${\cal A}_n$ on fuzzy torus ${\cal
T}$ as the Heisenberg Weyl group $Z_n \times Z_n$}

\indent

From section 2 we know that the effects of the noncommutative Wilson loop
\cite{AMNS, BA} $W_1 = U_1^{\frac{1}{n}}$ and $W_2 = (U_2^{\frac{1}{n}})$
on the $i$-th soliton is to translate its position, from $z_i$ to $(z_i +
\frac{1}{n} \tau - \delta_{in}\tau)$ and $(z_i + \frac{1}{n} -
\delta_{in})$ respectively, or equivalently shift all $z_i$ by
$\frac{1}{n}$ or $\frac{\tau}{n}$ mod the torus ${\cal T}$, furthermore
equivalent to shift the coordinate origin $u$ by $\frac{1}{n}$ of
$\frac{\tau}{n}$ in {\it oposite direction}. Substituting in
(\ref{basis}), we find
\begin{eqnarray}
\label{funcform-w1}
\hat{\bf V}_1 v_a(z_1, \cdots, z_n) & = & (\prod_{i = 1}^{n - 1}
T^{(i)}_{\frac{\tau}{n}}) T^{(n)}_{\frac{\tau}{n} - \tau}
v_a(z_1, \cdots, z_n)\\
\label{matrixform-w1}
& = & v_{a - 1}(z_1, \cdots, z_n) = W_1 v_a(z_1, \cdots, z_n),\\
\label{funcform-w2}
\hat{\bf V}_2 v_a(z_1, \cdots, z_n) & = & v_a(z_1 + \frac{1}{n},
\cdots, z_n + \frac{1}{n} - 1)\\
\label{matrixform-w2}
& = & (-1)^{n+1}e^{2\pi i \frac{a}{n}} v_a(z_1, \cdots, z_n) = W_2
v_a(z_1, \cdots, z_n),
\end{eqnarray}
where
\begin{equation}
T^{(i)}_{a \tau} f(z) \equiv e^{\pi i a^2 \tau + 2 \pi i a z_i}f(z_1,
\cdots, z_i + a \tau, \cdots, z_n).
\end{equation}

Using basis $v_a$ (\ref{basis}), we have transform the action of the N.C.
Wilson loop of ${\cal T}_n$ from the shift in functional space form
(\ref{funcform-w1}) and (\ref{funcform-w2}) into the matrix operator form
(\ref{matrixform-w1}) and (\ref{matrixform-w2}), i.e.
\begin{equation}
\label{comm-rule}
(\hat{\bf V}_1)_{a b} = \delta_{a + 1, b}; \hspace{.5cm}
(\hat{\bf V}_2)_{a b} = \delta_{a b} \omega^{a}, \hspace{.5cm}
\omega = e^{\frac{2\pi i}{n}} = e^{i \theta_n}
\end{equation}
As expected this $\hat{\bf V}_1$ and $\hat{\bf V}_2$ satisfy the relations
\begin{equation}
\hat{\bf V}_1^\dagger \hat{\bf V}_1 = \hat{\bf V}_2^\dagger \hat{\bf V}_2
= 1, \quad \hat{\bf V}_1^n = \hat{\bf V}_2^n = 1, \quad \hat{\bf V}_1
\hat{\bf V}_2 = \hat{\bf V}_2 \hat{\bf V}_1 e^{2\pi i\theta_n},
\end{equation}
here $\theta_n = \frac{1}{n}$ is the noncommutative parameter for the $n$
2-branes \cite{SW}.

Remark: There are some subtle and crucial points in our paper which
usually cause confusion. It seem worth to be stressed and clarified here.
Originally for ${\cal H}_{\cal T}$ on total ${\cal T}$ we have the
commuting $U_1$ and $U_2$ with $\theta = 2 \pi Z$. After orbifolding, it
is clear \cite{CDS, SW} that ${\cal H}_{\cal T} = V_n \otimes {\cal L}$ is
acted by $U_i = W_i \otimes \hat{\bf V}_i^{-1}$, where the matrices $W_i$
satisfy $W_1 W_2 = W_2 W_1 \omega$, while the difference operators
\begin{equation}
\hat{\bf V}_1^{-1} \hat{\bf V}_2^{-1} = \hat{\bf V}_2^{-1} \hat{\bf
V}_1^{-1} e^{\theta - \frac{2 \pi i}{n}} \equiv \hat{\bf V}_2^{-1}
\hat{\bf V}_1^{-1} e^{- 2 \pi i \theta_n}.
\end{equation}
Generally $W_i^\alpha$ is represented by $Z_n \times Z_n$ matrix on
trivial $su(n)$ bundle $V_n$, while $\hat{\bf V}_\alpha^{-1}$ is the
translation of the origin of ${\cal T}$: $u = 0 \longrightarrow u = -
\frac{\alpha_1 + \alpha_2 \tau}{n}$, correspondingly, to shift the center
of mass of $n$ solitons $\bar{z}_c = \frac{1}{n} \sum_{i = 1}^n z_i$ by
$\frac{\alpha_1 + \alpha_2 \tau}{n}$. If we restrict to consider the
${\cal H}_n = V_n$ on subtorus ${\cal T}_n$, then we may either realize
the $W_i^\alpha$ in operator form i.e. in terms of $Z_n \times Z_n$ matrix
(\ref{matrixform-w1}) (\ref{matrixform-w2}) or realize it in functional
formalism i.e. in terms of difference operators (\ref{funcform-w1})
(\ref{funcform-w2}). These two forms is related by the Weyl Moyal
transformation on torus. But it happens that the $\hat{\bf V}_\alpha$
($\bar{z}_c \longrightarrow \bar{z}_c + \frac{\alpha_1 + \alpha_2
\tau}{n}$) yield a matrix transform on our basis function, so it is
equivalent to the operator form of $W_\alpha$. Thus for the total torus
the $W_\alpha$ and $\hat{\bf V}_\alpha^{-1}$ cancels, $U_\alpha = W_\alpha
\otimes \hat{\bf V}_\alpha^{-1}$ are commutative, as the (6.54) in
\cite{SW}.

The functional and the operator forms are related by Weyl Moyal
transformations on torus (see Appendix B of \cite{GHS}), functions $v_a$
behaves as the coherent state functions on torus.

On the space ${\cal H}_n = \{ v_a | a = 1, 2, \cdots, n \}$, the {\it
noncommutative algebra ${\cal A}_n$ generated by $W_1$ and $W_2$} has been
truncated and becomes a $n \times n$ dimensional unital $C^*$ algebra with
$n^2$ basis $W_1^{\alpha_1} W_2^{\alpha_2} = W^{\alpha}$ and the define
relations
\begin{equation}
\label{w-comm}
W^{\alpha \beta}W^{\alpha \beta \dagger} = 1, \quad W^{\alpha
\beta} W^{\alpha^\prime \beta^\prime} = W^{\alpha^\prime \beta^\prime}
W^{\alpha \beta} \omega^{\alpha \beta^\prime - \beta \alpha^\prime} =
W^{\alpha + \alpha^\prime, \beta + \beta^\prime}\omega^{-\beta
\alpha^\prime}, \quad (W^{\alpha \beta})^n = 1.
\end{equation}
Here and hereafter to represent in matrix (operator) form the basis
$W^{\alpha_1, \alpha_2}$ of ${\cal A}_n$ we introduce the usual $Z_n
\times Z_n$ matrix $(I_\alpha)_{a b} = (I_{\alpha_1, \alpha_2})_{a b} =
\delta_{a + \alpha_1, \alpha_2}\omega^{b \alpha_2}$ \cite{Tata}. This
algebra in difference operator form is the Heisenberg group $G_{\cal
H}(n)$ $Z_n \times Z_n$ in ordinary $\theta$ function theory \cite{Tata},
including both the shift $\frac{\alpha \tau}{n} + \frac{\beta}{n}$ and the
change of phases.

\subsection{$su(n)$ algebra $g$ on torus}

\indent

It is shown \cite{CFH} that the level $l$ representation of the Lie
algebra $sl_n({\cal T})$ on the elliptic curve ${\cal T}$ can be written
as following:
\begin{equation}
\label{E_alpha}
E_{\alpha} = (-1)^{\alpha_1}\sigma_{\alpha}(0)\sum_j\prod_{k \neq j}
\frac{\sigma_{\alpha}(z_{jk})}{\sigma_0(z_{jk})}\left [ \frac{l}{n}
\sum_{i \neq j}\frac{\sigma_{\alpha}^\prime(z_{ji})}
{\sigma_{\alpha}(z_{ji})} - \partial_j \right ],
\end{equation}
and
\begin{equation}
\label{E_0}
E_0 = -\sum_j \partial_j,
\end{equation}
here $\alpha \equiv (\alpha_1, \alpha_2) \in Z_n \times Z_n$ and $\alpha
\neq (0, 0) \equiv (n, n)$, $z_{jk} = z_j - z_k$, $\partial_j =
\frac{\partial}{\partial z_j}$. $E_0$ commutes with $E_{\alpha}$,
$sl_n({\cal T})$ includes only the $E_{\alpha}$ with $\alpha \neq 0$.
After a complicate calculation, we obtain the commutation relation:
\begin{equation}
\label{E-comm}
[E_{\alpha}, E_{\gamma}] = (\omega^{-\alpha_2 \gamma_1} -
\omega^{-\alpha_1 \gamma_2}) E_{\alpha + \gamma},
\end{equation}
or in usual $sl(n)$ with $i, j$ label the Chan Paton indices basis, let
\begin{equation}
\label{E_ij}
E_{ij} \equiv \sum_{\alpha \neq 0} (I^\alpha)_{ij} E_{\alpha},
\end{equation}
then
\begin{equation}
[E_{jk}, E_{lm}] = E_{jm}\delta_{kl} - E_{lk}\delta_{jm}.
\end{equation}
Remark: The representation (\ref{basis}) and the commuation rules
(\ref{E-comm}) can also be obtained \cite{CFH} from a quasiclassical limit
from the representation of the $Z_n \times Z_n$ Sklyanin albebra
\cite{Sklynin}. Sklyanin and Takebe \cite{ST1} give the elliptic
$sl(2)$ by using double periodic Weierstrass functions. The high spin
$l$ representations is given also. In this paper, it is restricted to
the $l = 1$ representation of $sl_n({\cal T})$ by holomorphic sections
on ${\cal T}^{\otimes n}/{S^n}$.

\subsubsection{ Automorphism of $E_{\beta} \in su_2({\cal T})$ by
noncommutative gauge transformation $W^{\alpha} \in {\cal A}$}

\indent

Since the Wilson loops $W_1$ and $W_2$ acting on the noncommutative
covering torus ${\cal T}$ is to shift $z_i$ to $(z_i + \frac{\tau}{n} -
\delta_{in}\tau)$ and $(z_i + \frac{1}{n} - \delta_{in})$ respectively, so
$E_{\alpha}$ in (\ref{basis}) will be changed into
\begin{eqnarray}
\label{wE_alpha}
W_1\cdot (E_\alpha(z_i)) & = & W_1 E_\alpha(z_i) W_1^{-1}  =
E_\alpha(z_i + \frac{\tau}{n} - \delta_{in}\tau) =
\omega^{-\alpha_2}E_\alpha(z_i),\nonumber\\
W_2\cdot (E_\alpha(z_i)) & = & W_2 E_\alpha(z_i) W_2^{-1} =
E_\alpha(z_i + \frac{1}{n} - \delta_{in}) =
\omega^{\alpha_1}E_\alpha(z_i).
\end{eqnarray}
or more generally
\begin{equation}
\label{automorphism}
W^{\beta_1, \beta_2}\cdot E_\alpha = W^{\beta_1, \beta_2} E_\alpha
(W^{\beta_1, \beta_2})^{-1} = \omega^{\alpha_1\beta_2 - \alpha_2\beta_1}
E_\alpha.
\end{equation}
This could be compared with the matrix model on noncommutative torus
\cite{CDS}, where
$$
U_i X_j U_i^{-1} = X_j + \delta_{ij} 2 \pi R_j
$$
or more exactly with the covariant derivatives
$$
U_i \bigtriangledown_j U_i^{-1} = \delta_{ij}\bigtriangledown_j
$$
i. e.
$$
E_\alpha \Longrightarrow \exp(\frac{i \pi}{n}
\bigtriangledown_1)^{\alpha_1} \exp(\frac{i \pi}{n}
\bigtriangledown_2)^{\alpha_2}.
$$

\subsubsection{Isomorphism of $su_n({\cal T})$ and ${\cal A}_n$ on
${\cal H}_n$}

\indent

let $E_\alpha \in g$  to act on $v_a$, we find that
\begin{equation}
\label{E-effect}
E_\alpha v_a = \sum_b(I_\alpha)_{b a}v_b.
\end{equation}
As in section 3.2 we already learn that $W^\alpha v_a = \sum
(I_\alpha)_{ba}v_b$, so on ${\bf\cal H}_n$ we establish the isomorphism
\begin{eqnarray}
\label{isomorphism}
su_n({\cal T}) & \Longrightarrow & {\bf\cal A}\nonumber\\
E_\alpha & \longrightarrow & W^\alpha
\end{eqnarray}

Obviously this is correspondent with the noncommutative plane case, where
one have
\begin{equation}
\label{f-corp}
i\epsilon_{ji}\partial_{x_i} F(x) \Longrightarrow [x, * f(x)]
\end{equation}
and acting on Fock space
\begin{equation}
\label{op-corp}
\partial_z \longrightarrow a, \hspace{1cm} \partial_{\bar{z}}
\longrightarrow a^\dagger
\end{equation}
Compare eq.(\ref{f-corp}) with eq.(\ref{automorphism}) and
eq.(\ref{op-corp}) with (\ref{E_alpha}) and (\ref{E-effect}), it is easy
to see that the infinitesimal translations on plane ${\bf R}^2$
corresponds to $su_n({\cal T})$ and the algebra ${\cal A}_{{\bf R}^2}$
correponds to the algebra ${\cal A}_n$. Meanwhile the adj operation for
${\cal A}_{{\bf R}^2}$ (\ref{f-corp}) changes to the Adj operation (gauge
transformation \cite{AMNS, BA}) for ${\cal A}_n$ (\ref{wE_alpha}).

\subsubsection{$\hat{V}_i$ and $E_\alpha$ as generators of the Weyl
reflection group}

\indent

Kac \cite{Kac} has shown that the $\theta$ function with characteristic
$\alpha = (\alpha_1, \alpha_2) \in Z_n \otimes Z_n$ transforms under
affine Wely reflections as the Heisenberg group generated by $\hat{V}_i$
(\ref{funcform-w1}), (\ref{funcform-w2}). As we have shown in
(\ref{matrixform-w1}), (\ref{matrixform-w2}) and (\ref{E-effect}), both
$\hat{V}_\alpha$ and $E_\alpha$ act on $v_a$ as $I_\alpha$, operators
$\hat{V}_1$ and $E_1$ as the coxeter element $I_1$, operators $\hat{V}_2$
and $E_2$ as the cyclic element $I_2$; operator $E_{i j} (i \neq j)$
permutes $i, j$ i.e. gives the reflection $\hat{r}_{i j}$ reversing the
root $e_i - e_j$.

Here, let us stress the relations of N.C. algebras ${\cal H}_n$, ${\cal
A}_n$ and $su(n)$ with the solitons. Obviously, since in (\ref{basis})
each term $F_\alpha$ are symmetrical with respect to permutation $S_n$ of
the centers $z_i$ of solitons, so the vectors $v_\alpha$ is defined on the
module ${\cal T}_n^{\otimes n}/S_n$. The operator $\hat{\bf V}_1$ shifts
each soliton $z_i$ from the $i$-th covering of ${\cal T}$ over ${\cal
T}_n$ i.e. from the $i$-th brane to the next one, or equivalently changes
each brane to the previous one. The operator $\hat{\bf V}_2$ shifts the
$U(1)^{\otimes n}$ phase of $i$-th brane by $\omega^i$. The Chan Paton
$su(n)$ of $n$ branes is generated by $E_{i j}$. Acted on ${\cal H}_n$,
both $\hat{\bf V}_i$ and $E_{i j}$ generate the Weyl reflection. The
offdiagonal elements give permutations of the branes, the diagonal
elements generates the phase shift of soliton wave functions.

\subsubsection{Projection operators and soliton generating ABS operators}

\indent

In the operator formalism, the projection operators becomes
\begin{equation}
\frac{1}{n}\sum_\beta W^{0 \beta} (I_\beta)_{i i} = P_i =
\frac{| V_i \rangle \langle V_i |}{\langle V_i | V_i \rangle},
\end{equation}
and the ABS operators is simply
\begin{equation}
E_{1 0} \cong W_1 = \sum_a \frac{|V_{a + 1}\rangle\langle V_a |}{(\langle
V_{a+1} | V_{a + 1} \rangle \langle V_a | V_a \rangle)^{\frac{1}{2}}}.
\end{equation}
The explicit functional expresions may be obtained by Weyl Moyal
transformation adopted to the torus as described in the Appendix B of
\cite{GHS}.

\section{Affine algebra on N.C. torus, integrable elliptic Gaudin \\
and Calogero Moser models \cite{CFH}}

\indent

Now we will show that the cotangent bundle for N.C. elliptic $su_n({\cal
T})$ (\ref{E_alpha}) with twisted loop $W_i$ realizes the elliptic Gaudin
model on N.C. torus, which is related to critical level twisted $su(n)$
WZW i.e. $A_{n - 1}$ affine algebra on N.C. torus. Then use the elliptic
Vandermonde determinant to gauge transform it into elliptic C.M. model
which is the Hamiltonian reduction of the cotangent bundle for the algebra
of the semidirect product of the Cartan torus and the Weyl group
\cite{HM}.

\subsection{N.C. Elliptic Gaudin Model}

\indent

The elliptic Gaudin model on commutative space \cite{ST1} is defined by
the transfer matrix (quantum lax operator):
\begin{equation}
\label{L(u)}
L_{i j}(u) = \sum_{\alpha \neq (0, 0)} w_\alpha (u) E_\alpha(I_\alpha)_{i
j},
\end{equation}
where
\begin{equation}
\label{w_alpha(u)}
w_\alpha(u) = \frac{\theta^\prime(0) \sigma_\alpha(u)}
{\sigma_\alpha(0) \sigma_0 (u)},
\end{equation}
$E_\alpha$ and $I_\alpha$ are the generators of $su(n)$ (or ${\cal
A}_{n-1}$ Weyl) and $G_{\bf\cal H}(n)$ respectively. Using the general
defining relations of $su(n)$ (\ref{E-comm}) and $G_{\cal H}(n)$
(\ref{w-comm}), we find that the commutators of $L$ (quantum version of
fundamental Poisson bracket):
\begin{equation}
\label{L-comm}
[L^{(1)}(u_1), L^{(2)}(u_2)] = [r^{(1, 2)}(u_1, u_2), L^{(1)}(u_1) \oplus
L^{(2)}(u_2)],
\end{equation}
where the classical Yang-Baxter matrix
\begin{equation}
\label{r(u)}
r(u)_{i, j}^{k, l} = \sum_{\alpha \neq (0, 0)}w_\alpha(u) (I_\alpha)_i^k
\otimes (I_\alpha^{-1})_j^l
\end{equation}
is antisymmetrical
\begin{equation}
r_{i, j}^{k, l}(u) = r_{j, i}^{l, k}(-u)
\end{equation}
and satisfies the classical Yang-Baxter equation:
\begin{equation}
[r_{1 2}(u_{1 2}), r_{1 3}(u_{1 3})] + [r_{1 2}(u_{1 2}), r_{2 3}(u_{2 3}]
+ [r_{1 3}(u_{1 3}), r_{2 3}(u_{2 3}] = 0,
\end{equation}
where $u_{i j} = u_i - u_j$. Thus this system is integrable. Indeed, we
can find from the quantum determinant of $L$ the complete set of
Hamiltonian. Some examples may be found in \cite{ST1, ER, KTa}. It was
obtained also as the nonrelativistic limit of the Ruijsenaars-Macdonald
operators which will be described in next section. As a lattice model, the
common eigenfunctions and eigenvalues of Gaudin model is solved in terms
of Bethe ansatz \cite{FFR}, which has been expressed by the conformal
blocks of the twisted WZW models on the torus \cite{FW, KT}.

To relate these well known results about this usual Gaudin model with that
on the fuzzy torus ${\cal T}_n$ (section 3), let us substitute the
differencial representation (\ref{E_alpha}) of $su_n({\cal T})$ $E_\alpha$
into the Gaudin $L$ (\ref{L(u)}), then it turns to be in a factorized form
\cite{CFH}:
\begin{equation}
\label{factorized_L}
E_0 + \sum_{\alpha \neq (0, 0)}w_\alpha(u) E_\alpha (I_\alpha)_j^i =
L(u)_j^i = \sum_k \phi(u, z)_k^i \phi^{-1}(u, z)_j^k \partial_u - l \sum_k
\partial_u \phi(u, z)_k^i \phi^{-1}(u, z)_j^k ,
\end{equation}
where the factors are the vertex face intertwiner
\begin{equation}
\label{intertwinner-face}
\phi(u, z)_j^i = \theta_{\frac{1}{2} - \frac{i}{n}, \frac{1}{2}}(u + n z_j
- \sum_{k = 1}^n z_k + \frac{n-1}{2}, n \tau).
\end{equation}
We can also show that (\ref{factorized_L}) is a representation of $L$ by
subsituting it directly into (\ref{L-comm}). Now the expression of $r$
becomes
\begin{equation}
\label{intertwinner-factor}
r_{i, j}^{l, k}(u) = \delta_{i + j}^{l + k} \left \{ (1 - \delta_i^l)
\frac{\theta^0(0)^\prime \theta^{i - j}(u)}{\theta^{l - j}(u) \theta^{i -
l}(u)} + \delta_i^l \left ( \frac{\theta^{i - j}(u)^\prime}{\theta^{i -
j}(u)} - \frac{\theta(u)^\prime}{\theta(u)} \right ) \right \},
\end{equation}
where $\theta^i(u) \equiv \theta_{\frac{1}{2} - \frac{i}{n},
\frac{1}{2}}(u, n \tau)$. It is easy to find that it is $Z_n \otimes Z_n$
symmetrical $r(u) = (I_\alpha \times I_\alpha) r(u) (I_\alpha^{-1} \otimes
I_\alpha^{-1})$ and turns to be equal to (\ref{r(u)}), i.e. actually the
factorized $L_j^i$ (\ref{factorized_L}) represent (\ref{L-comm}).

The intertwinner "factor"s (\ref{intertwinner-face}) , now intertwines
the Chan Paton $su(n)$ index $i$ for the brane (vertex model index) with
the dynamical indices (face model indices) of dynamical soliton position
$z_j$ on the world sheet.

Feigin et. al \cite{FFR} has established the relation between the critical
level $su(n)$ WZW models and the rational Gaudin model, that is the $L(u)$
and $R(u)$ as (\ref{L(u)}) and (\ref{r(u)}), but instead of
(\ref{w_alpha(u)}) now with a rational $w_\alpha^R(u) = \frac{1}{u}$, $u
\in {\bf C}$. Kuroki and Takebe \cite{KTa} find the same relation for
the elliptic case, but with the WZW twisted, i.e. defined on a twisted
bundle $g^{tw}$:
$$
g^{tw} := ({\bf C} \times g)/\sim
$$
where the equivalence relation $\sim$ are $(u, g) \sim (u + 1, I_1 g
I_1^{-1}) \sim (u + \tau, I_2 g I_2^{-1})$. Correspondingly the
$w_\alpha(u)$ (\ref{w_alpha(u)}) is a meromorphic section of the twisted
bundle with single pole at $u = 0$ and residue $1$. While wrapping around
different one cycle on the base torus the global gauge transformation for
sections in the fibre are noncommute, but the base torus itself has the
complex conformal structures only, it has neither metric nor symplectic
structure, thus remains to be commutative. For our N.C. torus we have both
$z = y_1 + i y_2$ and $\bar{z} = y_1 - i y_2$ which are N.C., it causes
both the global N.C. $W_1, W_2$ and the local N.C. $z, \bar{z}$.

For the usual Gaudin \cite{ST1, ER} on a marked commutative torus, the
$z_i$ is the marked points, i.e. the poles in $L = \sum_i \sum_\alpha
w_\alpha(u - z_i) E_\alpha I_\alpha$. The $p_i$ is the conjugate momentum:
classically $\{p_i, q_j\} = \delta_{i j}$, quantumly $[p_i, q_j] =
\delta_{i j}$. So, the $L$-matrix is endowed with dynamics, by finding
Poisson (equivalently Konstant-Kirrilor) brackets at first, then by
quantized the $p_i \sim -i \frac{\partial}{\partial q_i}$ to find the
matrix differential form of $L$ \cite{CFH}. While now for the Gaudin on
noncommutative torus, the $z_i$ are the center (position) of the $i$-th
soliton, $\partial_i$ as its infinitesimal translation is equivalently to
the $[z_i *, ]$. Here we should stress that the key point to translate the
usual integrable models on commutative  space to that of N.C. space is to
notice that since in N.C. space  $[\bar{z}_i, z_j] = \delta_{i j}$, so
N.C. plane automatically become the symplectic manifold corresponding to
the quantum phase space. $p_i \rightarrow \bar{z}_i$, $q_i \rightarrow
z_i$ i.e in the holomorphic Fock Bargmann formalism $[\bar{z}, f]$
automatically becomes $\partial_z f$ just like $p \rightarrow
\frac{\partial}{\partial q}$. Apparently the complex structure with given
metric gives the symplectic structure, moreover it is automatically
"quantized", as the geometrical quantization.

Let us show that the dynamical equation induced by N.C. this way really
gives the interaction between N.C. solitons \cite{GHS}, by studying the
potential of C.M. model.

\subsection{Elliptic Calogero Moser model and its equivalence with
elliptic Gaudin model}

\indent

The elliptic C.M. model is defined by the Himiltonian:
\begin{equation}
H = \sum_{i = 1}^n \partial_i^2 + \sum_{i \neq j} g \wp(z_{i j}),
\end{equation}
where $\wp(z) = \partial^2 \sigma(z)$. This quadratic and other higher
Hamiltonian are generated by the Krichver Lax matrix
\begin{equation}
L_{\rm Kr}(u)_j^i = \partial_i \delta_j^i + (1 - \delta_j^i) \sqrt{g}
\frac{\sigma(u + z_{j i})}{\sigma(u) \sigma(z_{j i})}.
\end{equation}
By the Poisson transformation (classically $\partial_i \sim p_i$, $z_i
\sim q_i$)
\begin{equation}
p_i \longrightarrow p_i - \frac{\partial}{\partial q_i} \ln
\Pi^{\frac{l}{n}}(q),
\end{equation}
here
\begin{equation}
\Pi(q) \equiv \prod_{i < j}\sigma(q_{i j}), \hspace{1cm} \sqrt{g} = -
\frac{l}{n} \sigma^\prime(0),
\end{equation}
and $L_{\rm Kr}(u)_j^i$ becomes
\begin{equation}
\label{L_CM}
L_{\rm CM}(u)_j^i = (p_i - \frac{l}{n}\frac{\partial}{\partial q_i}\ln
\Pi(q))\delta_j^i - \frac{l}{n}\sigma^\prime(0)(1 - \delta_j^i)
\frac{\sigma(u + q_{j i})}{\sigma(u) \sigma(q_{j i})}.
\end{equation}
This may be further gauge transformed into the factorized $L$
(\ref{factorized_L}) of Gaudin model by gauge transformation matrix
\begin{equation}
G(u; q)_j^i \equiv \frac{\phi(u; q)_i^i}{\prod_{l \neq
j}\sigma(q_{j l})}.
\end{equation}

It is well known that C.M. model gives the dynamics of a long distance
interaction between $n$-bodies located at $z_i$, $(i = 1, 2, \cdots, n)$.
Now we will show that it is the interaction between $n$-solitons on the
fuzzy torus while $z_i$ becomes the positions of the centers of each
soliton. According to paper \cite{GHS}, for N.C. multisolitons the
potential term is argumented to be the Laplacian of a K\"ahler potential
$K$, which is the logrithm of a Vandermonde determinant. Actually we have
\begin{eqnarray}
\sum_{i \neq j}\wp(z;j) & = & \sum_i \partial_i^2 \log \prod_{j \neq k}
\sigma(z_j - z_k) \equiv \sum_i \partial_i^2 K(u, z),\\
e^{K(u, z)} & = & \prod_{j \neq k} \sigma(z_j - z_k)\sigma(n u + \frac{n(n
- 1)}{2})\nonumber\\
& = & \det(\phi_k^j) \equiv \sigma(n u + \frac{n(n - 1)}{2})\prod_{i \neq
j}\sigma(z_{i j}).
\end{eqnarray}
The variable $u$ of the marked torus is the so called spectral parameter
or evaluation parameter of Lax matrix $L_{\rm Kr}(u)_j^i$.

As \cite{BCS}, replacing the "spin" index $i, j$ elements $E_{i j}$
(\ref{E_ij}) in the Lax matrix $L_{\rm Kr}(u)_j^i$ of the C.M. model by
soliton exchange $z_i \leftrightarrow z_j$ permutation $s_{i j}$ in ${\cal
A}_{n-1}$ Weyl reflection, $L_{\rm Kr}(u)_j^i$ becomes like the Dunkle
operators \cite{BFV}. This "spin" index exchange equivalence with particle
(soliton) exchange is obtained if restricted to be acted on the space of
wave functions which is totally symmetric in both spins and positions.
This is the ${\cal A}_{n-1}$ Weyl symmetry for the $su_n({\cal T})$ Chan
Paton index of the branes and for the positions $z_i, z_j$ of solitons.

Here the determinant of the vertex-face interaction is the determinant of
our gauge transformation matrix from Gaudin to C.M. We should mention that
its $u$ independent factor $\prod(z)$ is the Weyl antiinvariant ground
state wave function of C. M. It is the "phase functions" part of the
conform block of the twisted WZW model on elliptic curve related to the
Gaudin model \cite{FFR, FW, KT}. Dividing by this phase the KZB equation
of the elliptic Gaudin Model reduces \cite{FW, KT} to the heat equation
associated to the elliptic C. M. equation.

\section{The $Z_n \times Z_n$ Heisenberg group realized by the
Sklyanin algebra ${\bf\cal S}_{\tau, \eta}$ in case of general $\theta$}

\indent

In previous section, we have show the utility of ellipitc Gaudin-Calogero
model for the N.C. field, brane and QH fluid in case of rational $\theta$,
it is reasonable to further find some kind of N.C. solitons for the
generic, i.e. deformed $\theta$. Acturally, there is a lot of hints, that
the Ruijsenaars-Schneider (RS) model, as the deformed version of C.M.,
descrbes some brane configurations. The elliptic RS and the infinitely
related IRF etc. may be the hopeful candinate.

Firstly let us remind the physical meaning of unrational $\theta$. If we
fix the area of the base torus, but change the field $B \sim
\frac{1}{\Theta}$, such that the passed $B$ flux quantum $\frac{\theta}{2
\pi}$ ("normalized area") generally is not an integer, then the commutator
$U_1 U_2 U_1^{-1} U_2^{-2} = e^{i \theta}$ becomes generic, unrestricted
to the root of unity. This implies also that the ratio of the area for the
Narain dauls $SL(2Z) \otimes SL(2Z)$ (modular for base ${\cal T}$, Morita
for target) is generic, i.e. unrestricted to be rational. We conject that
it is related to the characteristical parameter $\eta$ of IRF model etc.
i.e. we assume that $\theta \equiv \eta \tau$ mod modular (Morita)
transformation.

Now instead of simply shift by $\frac{1}{n}$ and $\frac{\tau}{n}$ under
the loop action $W_i$ (\ref{funcform-w1}) and (\ref{funcform-w2}) or the
$su(n)$ covariant differencial operator $E_{\alpha}$, we introduce the
difference operator $S_{\alpha}$ \cite{Sklynin, HSY} as following
\begin{eqnarray}
\label{S_alphaF}
S_{\alpha} F & = & \sum_j S_{\alpha}^{(j)} F(z_1, z_2, \cdots, z_j - \eta,
z_{j + 1}, \cdots, u_n),
\end{eqnarray}
where
\begin{equation}
\label{S_alpha^j}
S_{\alpha}^{(j)} = e^{\pi i \alpha_2} \sigma_\alpha(\eta)\prod_{k \neq j}
\frac{n\sigma_\alpha(n z_j - n z_k + l \eta)}{\sigma_0(n \eta)\sigma_0(n
z_j - n z_k)}
\end{equation}
Remark: The commutative relation of define $S_\alpha$ is given in
\cite{Sklynin} for $n = 2$ and \cite{HSY} for general $n$. The difference
representation of $S_\alpha$ is constructed and proved via Belevin's
$R$-matrix by \cite{Sklynin, HSY} and then proved directly by \cite{HSY}
and \cite{HSY1}

\noindent{\bf Fundamental representation}

Applying $S_\alpha$ with level $l = 1$ on $V_a \in {\cal H}_n$
(\ref{basis}), we get
\begin{equation}
S_\alpha V_a = \sum_b (I_\alpha)_{ab} V_b.
\end{equation}
Actually, acting on ${\cal H}_n$, the fundamental representation of
$S_\alpha \in {\cal S}_{\tau, \eta}$ becomes the Heisenberg $Z_n \times
Z_n$ matrix $I_\alpha$. We prove this by three lemmas.

\noindent {\bf Lemma 1}. The $S_0 F_0$ is independent of $\eta$,
\begin{equation}
\label{lemma1}
S_0 F_0 = F_0.
\end{equation}
From (\ref{S_alphaF}), (\ref{S_alpha^j}) and (\ref{basis}) we get
\begin{eqnarray}
\label{sigmaSF}
\sigma_0(n \eta) S_0 F_0 & = & n\sigma_0(\eta)\sum_j \left [ \prod_{k \neq
j}\frac{\sigma_0(n z_j - n z_k + \eta)}{\sigma_0(n z_j - n z_k)} \right ]
\nonumber\\
& & \cdot \sigma_0(n z_j - \sum z_k - (n - 1)\eta)\prod_{l \neq
j}\sigma_0(n z_l - \sum z_k + \eta).
\end{eqnarray}
As a function of $\eta$ with given $z_i$ we find that the right hand side
(r.h.s) of (\ref{sigmaSF}) has $n^2$ zeros at $\eta = \frac{1}{n}(\alpha_1
\tau + \alpha_2)$.

In the quasiperiodic phase of the r.h.s under shift $\eta \longrightarrow
\eta + \tau$, the $\eta$ dependent part is $\exp(- 2 \pi i n^2 \eta)$,
that is the r.h.s has exactly $n^2$ zeros as the function $\sigma_0(n
\eta)$. Thus $S_0 F_0$ is independent of $\eta$. Take $\eta \rightarrow 0$
limit we find (\ref{lemma1}).

\noindent {\bf Lemma 2.}
\begin{equation}
\label{lemma2}
S_{0 0} F_\gamma = F_\gamma.
\end{equation}
This can be obtained from lemma 1 by further applying Heisenberg shift
$z_1 = z_1 - \frac{1}{n}(\gamma_1 \tau + \gamma_2)$ on lemma 1 i.e. on
both side of (\ref{sigmaSF}) as following:
\begin{eqnarray}
\sigma_0(\eta) & \sum_j & [\prod_{k \neq j}\frac{\sigma_0(n z_j - n z_k
+ \eta)}{\sigma_0(n z_j - n z_k)}]\nonumber\\
& \cdot & \sigma_\gamma(n z_j - \sum z_k - (n - 1)\eta) \prod_{l \neq j}
\sigma_\gamma(n z_l - \sum z_k + \eta)\nonumber\\
& = & \frac{1}{n}\sigma_0(n \eta) \prod_j (n z_l - \sum z_k),
\end{eqnarray}
or simply
\begin{equation}
S_{0 0} F_\gamma = F_\gamma.
\end{equation}
In fact, it is independent of which $z_i$ only that is shift to $z_i -
\frac{1}{n}(\gamma_1 \tau + \gamma_2)$. Since in the original and shifted
$F_\alpha$, the $n z_j - \sum z_k$ are weight vectors of ${\cal A}_{n -
1}$, all will be shifted by $(\gamma_1, \gamma_2)$ in the coset (weight
lattice)/(root lattice). While in $S_{0 0}$ the root $n(z_j - z_k)$ with
$z_i$ (without $z_i$) changes (unchanges) by integer $(\gamma_1 \tau +
\gamma_2)$. So there appears only a phase shift for each $\sigma_0$, which
cancels in total. It is easy to see that this is equivalent to the
Heisenberg transformations (\ref{funcform-w1}), (\ref{funcform-w2}) and
(\ref{wE_alpha}) on ${\bf\cal H}_n$ where we have shift all the $z_i$ at
the same time to the next coverings etc.

\noindent {\bf Lemma 3}. Transform lemma 2 by
\begin{equation}
\label{transform}
\eta \longrightarrow \eta + \frac{1}{n}(\alpha_1 \tau + \alpha_2)
\end{equation}
we have
\begin{eqnarray}
\label{lemma31}
S_\alpha F_{\gamma + \alpha} & = & e^{2 \pi i \alpha_2(\gamma_1 +
\alpha_1)/n} F_\gamma,\nonumber\\
S_\alpha F_\beta & = & \omega^{\alpha_2 \beta_1} F_{\beta - \alpha}
\end{eqnarray}
Really substituting (\ref{transform}) into (\ref{lemma2}), we get
\begin{eqnarray}
\sigma_\alpha(\eta) & \sum_j & [\prod_{k \neq j}\frac{\sigma_\alpha(n z_j
- n z_k + \eta)}{\sigma_0(n z_j - n z_k)}]\nonumber\\
& \cdot & \sigma_{\gamma + \alpha}(n z_j - \sum z_k - (n - 1)\eta)
\prod_{l \neq j}\sigma_{\gamma + \alpha}(n z_l - \sum z_k +
\eta)\nonumber\\
& = & \frac{1}{n}\sigma_0(n \eta)e^{\frac{2 \pi i \alpha_2}{n}(\gamma_1 +
\alpha_1)}\prod_j \sigma_\gamma(n z_j - \sum z_k).
\end{eqnarray}

For higher representation, the fussion rule $l = l_1 + l_2$ for $S_\alpha$
has been given in \cite{HSYY}. The representation space basis has been
given by Z. X. Yang. The cyclic representation was given in \cite{HSY}

\noindent {\bf Heisenberg automorphism of ${\cal S}_{\tau, \eta}$}

By shifting $\eta \longrightarrow \eta + \frac{1}{n}(\alpha_1 \tau +
\alpha_2)$ and $z_1 \longrightarrow z_1 - \frac{1}{n}(\alpha_1 \tau +
\alpha_2)$, from lemma 1 and lemma 2 it is easy to check that there is an
automorphism:
\begin{equation}
S_\beta \longrightarrow S_\beta^\prime = S_{\beta + \alpha}
\omega^{\alpha_2 \beta_2 + \beta_1 \beta_2}, \hspace{1cm} F_\gamma
\longrightarrow F_\gamma^\prime = \omega^{\alpha_2 \gamma_1} F_\gamma
\end{equation}
which satisfy
\begin{equation}
S_\beta^\prime V_a^\prime = \sum_b (I_\beta)_{a b} V_b^\prime.
\end{equation}
Remark: The automorphism for commutation rule of $S_\alpha$ and similar
change in basis for general level $l$ representations have been given in
the paper \cite{HSYY}

\noindent {\bf ${\cal A}_n$ and $su_n({\cal T})$ vs. ${\cal S}_{\tau,
\eta}$, $su_n({\cal T}) \sim \lim_{\eta \rightarrow 0}{\bf\cal S}_{\tau,
\eta}$}

Let $\eta \rightarrow 0$, the $S_\alpha$ (\ref{S_alphaF}),
(\ref{S_alpha^j}) in ${\bf\cal S}_{\tau, \eta}$ degenerates into the
$E_\alpha$ (\ref{E_alpha}) and (\ref{E_0}) in $su_n({\cal T})$ in the
first order of $\eta$, $S_\alpha \longrightarrow E_\alpha$, $(\alpha \neq
0, 0)$ while $S_{0, 0} \longrightarrow E_{0 0}$ is the zeros order.

The $su_n(\cal T)$ is an ordinary Lie algebra with highest weight
representation. But defining relations of ${\cal S}_{\tau, \eta}$ are
quadratic, no linear sum of $S_\alpha$ lies on the center of ${\cal
S}_{\tau, \eta}$ and ${\cal S}_{\tau, \eta}$ has no highest weight
representation.

The $S_\alpha$ is gauge transformed by the $Z_n \times Z_n$ $G_{\cal
H}(n)$ shifts of $z_i$ among coverings, and meantime itself transforms
nonabelliantely from $S_\alpha$ to $S_{\alpha + \beta}$ by the $G_{\cal
H}(n)$ shift $\eta \longrightarrow \eta + \frac{1}{n}(\beta_1 + \beta_2
\tau)$. The $S_\alpha$ implicitely intertwin with the $Z_n \times Z_n$
symmetries. For example, in $(I^{\alpha \beta})_{i j}$, $i, j$ are the
indices of the solitons, $\alpha, \beta$ are the indices of branes
(coverings). Thus the affine Weyl reflections consist of the reflections
of solitons (in the center mass system) and the truncation by a period of
${\cal T}_n$ into another covering. So the operators $S_{\alpha_1
\alpha_2}$ interchange the wave functions from brane $\alpha_1$ to
$\alpha_2$. The change of the strenth $B$ of the fields; i.e. the strenth
$\eta$ by $\frac{\beta_1 + \beta_2 \tau}{n}$ induces the $su(n)$ gauge
transformation of the operators $S_\alpha$ to $S_{\alpha + \beta}$.

\noindent{\bf The $Z_n \otimes Z_n$ Belavin model}

Let
\begin{equation}
\label{L^B(u)}
L^{\rm B}(u)_{i j} = \sum_\alpha W_\alpha(u)(I_\alpha)_{i j} S_\alpha,
\end{equation}
where $W_\alpha(u) \equiv \frac{\sigma_\alpha(u +
\eta)}{\sigma_\alpha(\eta)}$. Then
\begin{equation}
R_{1 2}^{\rm B} (u_1 - u_2) L_1(u_1) L_2(u_2) = L_2(u_2) L_1(u_1) R_{1
2}^{\rm B}(u_1 - u_2),
\end{equation}
where $R^{\rm B}(u)^{j l}_{i k} = \sum_\alpha W_\alpha (I_\alpha)_i^j
(I^{-1}_\alpha)_k^l$ is the Belavin $Z_n \times Z_n$ vertex model. Let
$\eta \rightarrow 0$, then $R^{\rm B} \rightarrow R^{\rm G}$. Furthermore
following \cite{H1} using the same gauge transformation as we did from
$L^{\rm G} \rightarrow L^{|rm CM}$, one can obtain the Ruijsenaars
operator ${\cal H}_n$ as the $u$ independent part, this is very important
to understand the dynamics of soliton. And the $\det |L^{\rm B}(u) - k|$
will give the Ruijsenaars ${\cal H}_n$ at the $n$-th power of $k$: ${\rm
det}|L^{\rm B}(u) - k| = 0$ gives the spectral curve. All this will be
related to the elliptic quantum group $E_{\tau, \eta}$ and to be discussed
in the next section. And its relation to the N.C. field theory such as
the N.C. CS and gauged WZW will be investigated in the future.

\section{Elliptic quantum group and IRF model}

\indent

To define the {\it elliptic quantum group} $E_{\tau, \eta}(sl_2)$, we
start from a $n \times n$ matrix $R(u, \lambda) \in End(V \otimes V)$,
where $V$ is the element in representation space of $sl(n)$ and $\lambda$
a weight vector $\lambda = (\lambda_1, \lambda_2, \cdots, \lambda_n) \in
h^*$.
\begin{equation}
R(u, \lambda) = \sum_{i = 1}^n E_{i, i} \otimes E_{i, i} + \sum_{i \neq j}
\alpha(u, \lambda_{i j}) E_{i, i} \otimes E_{j, j} + \sum_{i \neq j}
\beta(u, \lambda_{i j}) E_{i, j} \otimes E_{j, i},
\end{equation}
where $E_{i, j}{\bf e}_k = \delta_{j, k}{\bf e}_i$, ${\bf e}_k = (0,
\cdots, 1, \cdots, 0)$, here $1$ is at the $k$-th place, and
$$
\alpha(u, \lambda) = \frac{\theta(u)\theta(\lambda + \eta)}{\theta(u -
\eta)\theta(\lambda)}, \hspace{1cm} \beta(u, \lambda) =
\frac{\theta(u + \lambda)\theta(\eta)}{\theta(u - \eta)\theta(\lambda)}.
$$
It satisfies the dynamical YBE:
\begin{eqnarray}
& &R(u_1, u_2, \lambda - \eta h^{(3)})^{12}R(u_1, \lambda)^{13} R(u_2,
\lambda - \eta h^{(1)})^{23}\nonumber\\
& & = R(u_2, \lambda)^{23} R(u_1, \lambda - \eta h^{(2)})^{13} R(u_1 -
u_2, \lambda)^{12}
\end{eqnarray}
where $R(u, \lambda - \eta h^{(3)})^{12}$ acts on a tensor $v_1 \otimes
v_2 \otimes v_3$ as $R(u, \lambda - \eta \mu) \otimes Id$ if $v_3$ has
weight $\mu$.

The elliptic Quantum group $E_{\tau, \eta}(sl_n)$ is an algebra generated
by a meromorphic function of a variable $h$ and the matrix elements of a
matrix $L(z, \lambda) \in End(\tilde{V})$ with noncommutative entries,
subject to the following relation
\begin{eqnarray}
\label{YBR}
&& R(u_1 - u_2, \lambda - \eta h^{(3)})^{12} L(u_1, \lambda)^{13} L(u_2,
\lambda - \eta h^{(1)})^{23}\nonumber\\
&& = L(u_2, \lambda)^{23} L(u_1, \lambda - \eta h^{(2)})^{13} R(u_1 - u_2,
\lambda)^{12}.
\end{eqnarray}

On the other hand, there is the well-known {\it interaction-round-a-face
(IRF) model}. We denote by $W(a, b, c, d, u)$ its Boltzmann weight for a
state configuration $\left ( \begin{array}{cc}
a & b\\
c & d
\end{array} \right )$ round a face with $a, b, c, d \in h^*$ which is the
weight space of the Lie algebra $sl_n$. The only nonvanishing Boltzmann
weights \cite{Quano} can be represented as
\begin{eqnarray}
\label{weights}
&& W(m, m - {\bf e}_j, m - 2 {\bf e}_j, m - {\bf e}_j, u) = \frac{\theta(u
+ \eta^\prime)}{\theta(\eta^\prime)}, \nonumber\\
&& W(m, m - {\bf e}_j, m - {\bf e}_j - {\bf e}_k, m - {\bf e}_j, u) =
\frac{\theta(u + \eta^\prime m_{j k})}{\theta(\eta^\prime m_{j k})},
\hspace{.5cm} (j \neq k)\\
&& W(m, m - {\bf e}_j, m - {\bf e}_j - {\bf e}_k, m - {\bf e}_k, u) =
\frac{\theta(u)\theta(\eta^\prime m_{j k} - \eta^\prime)}
{\theta(\eta^\prime)\theta(\eta^\prime m_{j k})}\nonumber
\end{eqnarray}
where $\eta^\prime = - \eta$, $m_{j k} = m_j - m_k$. They satisfy the YBE:
\begin{eqnarray}
\label{WYBE}
&& \sum_g W(a, b, c, g, u_1 - u_2) W(g, c, d, e, u_1) W(a, g, e, f,
u_2)\nonumber\\
&& = \sum_g W(b, c, d, g, u_2) W(a, b, g, f, u_1) W(f, g, d, e, u_1 -
u_2).
\end{eqnarray}

The $R$-operator as $n \times n$ matrix in auxiliary spaces becomes the
Boltzmann weight of the IRF model $W(a, b, c, d, z)$:
\begin{equation}
\frac{\theta(u - \eta)}{- \theta(\eta)} R(u, \eta m) {\bf e}_j \otimes
{\bf e}_k = \sum_{i, l} W(m, m - {\bf e}_j, m - {\bf e}_j - {\bf e}_k, m -
{\bf e}_i, u){\bf e}_i \otimes {\bf e}_l,
\end{equation}
where ${\bf e}_i + {\bf e}_l = {\bf e}_j + {\bf e}_k$. Thus, eq.
(\ref{YBR}) can be rewritten as
\begin{eqnarray}
\label{WYBR}
&&\sum_{j, j^\prime} W(a, b, c, d, u_1 - u_2) L^{(1)}(u_1, \lambda)^j_k
L^{(2)}(u_2, \lambda - k \eta)^{j^\prime}_{k^\prime}\nonumber\\
&& = \sum_{i_1, j_1} L^{(2)}(u_2, \lambda)^{l^\prime}_{j_1} L^{(1)}(u_1,
\lambda - j_1 \eta)^l_{i_1} W(a^\prime, b^\prime, c^\prime, d^\prime, u_1
- u_2),
\end{eqnarray}
where
\begin{eqnarray}
&& \eta a = \lambda - \eta h^{(3)} = \lambda - \eta \mu, \hspace{1cm} \eta
a^\prime = \lambda, \hspace{1cm} \eta^\prime = - \eta, \\
&& a - b = {\bf e}_j, \hspace{.5cm} b - c = {\bf e}_{j^\prime},
\hspace{.5cm} a - d = {\bf e}_{l^\prime}, \hspace{.5cm} d - c = {\bf
e}_l,\\
&& a^\prime - b^\prime = {\bf e}_k, \hspace{.5cm} b^\prime - c^\prime =
{\bf e}_{k^\prime}, \hspace{.5cm} a^\prime - d^\prime = {\bf e}_{j_1},
\hspace{.5cm} d^\prime - c^\prime = {\bf e}_{i_1},\\
&& {\bf e}_j + {\bf e}_{j^\prime} = {\bf e}_l + {\bf e}_{l^\prime},
\hspace{1cm} {\bf e}_k + {\bf e}_{k^\prime} = {\bf e}_{i_1} + {\bf
e}_{j_1}.
\end{eqnarray}
We will drop the superscripts $3$ of the quantum spaces later.

Compare (\ref{WYBE}) and (\ref{WYBR}), using explicite (\ref{weights}), it
is easy to give an evaluation representation of the elliptic quantum group
\begin{equation}
\label{L-intertwinner}
L(u, \lambda)^j_k = \frac{\sigma_0(u + \frac{\xi}{n} - \eta \delta - \eta
a_{k j} - \frac{n - 1}{2})}{\sigma_0(u - \eta \delta - \frac{n - 1}{2})}
\prod_{i \neq j}\frac{\sigma_0(-\frac{\xi}{n} + \eta a_{k
i})}{\sigma_0(\eta a_{j i})}
\end{equation}
which can be factorized as \cite{YHM}
\begin{equation}
\label{intertwinner}
L^{(1)}(u, \lambda) = \sum_i \bar{\phi}(u)^{a i}_{a -{\bf e}_j} \phi(u
+ \xi)^{a^\prime i}_{a^\prime - {\bf e}_k},
\end{equation}
where
$$
\phi^{a i}_b(u) = \left \{ \begin{array}{l}
\theta^{a i}_b (u + n \lambda_k - \sum \lambda_j, n \tau), \hspace{1cm} a
- b = {\bf e}_k,\\
0, \hspace{1cm}otherwise
\end{array} \right .
$$
and $\delta = \sum_{i = 0}^{n - 1} a_i$, $ a = \eta^{-1}(\lambda -
\eta h)$.

The evaluation Verma module $V_\Lambda(u)$ of $E_{\tau, \eta}(sl_n)$ is
acted by the $L(u, \lambda)^j_k$ with $\xi = \Lambda$, and a $f(h)$,
$$
f(h) L^j_k = L^j_k f(h - (j - k)).
$$
The basis ${\bf e}_k$ of $V_\Lambda(u)$ satisfies
$$
f(h) {\bf e}_k = f(\Lambda - k) {\bf e}_k, \hspace{.5cm}(k \geq 0);
\hspace{1cm} L_k^j {\bf e}_0 = 0, \hspace{.5cm} (j < k).
$$
If $\Lambda + 1 = n + \frac{\alpha + \beta \tau}{\eta}$, then the
vector ${\bf e}_0$ is the highest one in $V_\Lambda$, the subspace
of $V_\Lambda(u)$ spanned by ${\bf e}_k$, $(k > n - 1)$ is a submodule.
The quotient space $L_\Lambda(u)$ is a module of dim $n$.

Following \cite{FV2}, for an $E_{\tau, \eta}(sl_n)$ module $V$, let $F_{u
n}(V)$ be the space of meromorphic functions of $\lambda$ with values in
$V$ and $1$ periodic, $F(\lambda + 1) = F(\lambda)$. Introduce
indomorphism $f(\tilde{h})$, $\tilde{L}_k^j(w) \in End(F_{un}(V))$ by the
rule
\begin{equation}
(f(\tilde{h})F)(\lambda) = f(h) F(\lambda),
\end{equation}
and the difference matrix operators
\begin{equation}
(\tilde{L}_k^j(w) F)(\lambda) = L_k^j(w) F(\lambda - (j - k)\eta).
\end{equation}
This defines a difference matrix operator algebra $A(V)$ on $V$.

\noindent {\bf Factorizations of vertex model and IRF model}

Substituting the $S_\alpha(z)$ (\ref{S_alpha^j}) into the Belavin's
$L^B(u)_i^j$ (\ref{L^B(u)}), we find \cite{HSY} it can be factorized as
$\sum_a \bar{\phi}(u + \xi, z)_i^a \phi(u, z)_a^j e^{i\eta
\frac{\partial}{\partial z_a}}$. Thus, the intertwinner in
(\ref{intertwinner}) $\phi_a^i \sim \phi_{\lambda - {\bf e}_a}^{\lambda
i}$ intertwines the Heisenberg shift of $\frac{\tau}{n}$ $i$ in $\theta
\left
[ \begin{array}{c}
\frac{i}{n} + \frac{1}{2}\\
\frac{1}{2}
\end{array} \right ]$ for the $i$-th brane and the change of $\eta
\lambda_a$ ($\eta z_a$) by $\eta$ times unit weight vector $(z_a -
\frac{1}{n}\sum_k z_k)\eta$ for the $a$-th soliton.

The Heisenberg shift corresponds to the $W_\alpha$ matrix for wrap on
torus ${\cal T}$, while the later correspond to the difference of
$\lambda$ ($ \sim z$) in sections of bundle ${\cal L}$.

\indent

\noindent {\bf Transfer matrix and Ruijsenaars operators}

The {\it transfer matrix} is a difference operator acting on the space
$F(V[0])$  of meromorphic functions of $\lambda \in h^*$ with values in
zero weight space $V[0]$ of the quantum space ($h$ module) $V$. It is
defined by the formula
\begin{equation}
\label{transfer}
T(u)f(\lambda) = \sum_{i = 1}^n L_{i i}(u, \lambda)f(\lambda - \eta h).
\end{equation}

On the otherhand, as shown by Hasegawa by expanding the normal ordering
$\det|L^B(u) - t|$ \cite{H1}, one obtain the {\it Ruijsenaars Macdonald
operator $M$}
\begin{equation}
\label{M}
M = \sum_i^N \prod_{j:j \neq i} \frac{\theta(\lambda_i - \lambda_j +
l\eta)}{\theta(\lambda_i - \lambda_j)} T_i,
\end{equation}
and the other higher Hamiltonian.

The Ruijsenaars operators (\ref{M}) are integrable difference
operators that give a $q$- deformation of the Calogero-Moser
integrable differential operators (\ref{L_CM}), inversely
Calogero-Moser is the $\eta \rightarrow 0$ limit of
Ruijsenaars-Schneider.

Let us as \cite{EK} consider the transfer matrix associated to the
symmetric power $S^m V$. The zero weight space of this module is trivial
unless $m$ is a multiple of $n$. If $m = n l$ then the zero weight space
is one-dimensional and is spanned by the sum of the tensors $e_{i_1}
\otimes \cdots \otimes e_{i_{n l}}$ over all sequences $(i_j)$ such that
each number between $1$ and $n$ occurs precisely $l$ times. Let us denote
this sum by $e$.

Let us as \cite{FV} identify $S^{n l}(C^N)[0]$ using the basis $e$, and
let $T(z)$ be the transfer matrix associated to $S^{n l}V(0)$. Then
\begin{equation}
T(u) = \frac{\theta(u - \eta l)}{\theta(u - \eta n l)}M
\end{equation}
which is the same as from \cite{YHM} by choosing the parameters $\delta$,
$\xi$ appropriately.

Let us consider the $l = 1$ case, the $\lambda_i \sim z_i$ is the position
of our solitons. All the $M$ and other higher Hamiltonian depends only on
$z_{i j}$, and independent of neither spectral $u$ nor the center $\sum
z_i$. So the long distance interactions, the dynamics of solitons are
given by the elliptic Ruijsenaars-Schneider models.

\indent

\noindent {\bf Bethe Ansatz for the Hilbert space of non-commutative
torus}

Now we turn to find the common eigenvectors of the transfer matrix and the
Macdnoald operator $M$ (\ref{M}). It is easy to check that the vector
$v_a$ (\ref{basis}) in finite dimensional space $V_n$ are eigenvectors of
$M$ \cite{HSY1}. But the vector space $V_n$ as the trivial $U(n)$ boudle
\cite{SW} transformed only by $I^{\alpha \beta}$ matrix, representing the
$W_i$, independent of the twist $\eta$.

The noncommutative algebra ${\cal A}_\eta(n)$ for the twisted case should
be acted on the Hilbert space ${\cal H}_\eta(n) \sim {\cal L}\otimes V_n$
constituted by vector-valued functions rather than on finite dimensional
vector space $V_n$. The contribution of paper \cite{FV2} on the algebraic
Bethe ansatz for the elliptic quantum group is just to use transfer
matrices acted on spaces ${\cal H}_n$ of vector-valued functions.

For the module \cite{FV1} $L_\Lambda(u)$ ($\equiv {\cal H}_n$) (we
restricted this irre. case), the transfer matrices $T(u)$ preserves the
space $F_{u n}({\cal H})[0]$ of functions with values in the zero-weight
space ${\cal H}[0]$ and commute pair wise on it. The highest weight vector
of ${\cal H}$ is a $v \in F_{u n}({\cal H})(\Lambda)$ such that $L_j^i(u)
v = 0$ $(j > i)$, and $F_{u n}({\cal H})$ is spanned by the vector of the
form
\begin{equation}
\label{eigenvector}
L_{i_1}^{j_1}(t_1) \cdots L_{i_m}^{j_m}(t_m)v
\end{equation}
with all $j_m > i_m$.

Its eigenfunction is similar to that for the elliptic Gaudin. Now the
Hermite Bethe parameters satisfy the so called nested Bethe ansatz
equation \cite{HYZ, HKo}
\begin{eqnarray}
\label{BAE}
e^{\zeta_\alpha \eta} \prod_{l = 1}^{m_\alpha}\frac{\theta(t_j^{(\alpha +
1)} - t_l^{(\alpha)} - \eta)}{\theta(t_j^{(\alpha + 1)} - t_j^{(\alpha)})}
= \prod_{k = 1, k \neq j}^{m_{\alpha + 1}} \frac{\theta(t_j^{(\alpha +
1)} - t_k^{(\alpha + 1)} - \eta)}{\theta(t_j^{(\alpha + 1)} -
t_k^{(\alpha + 1)} + \eta)} \prod_{l = 1}^{m_{\alpha + 2}}
\frac{\theta(t_l^{(\alpha + 2)} - t_j^{(\alpha + 1)} -
\eta)}{\theta(t_l^{(\alpha + 2)} - t_j^{(\alpha + 1)})}.
\end{eqnarray}

The Bethe ansatz equation (BAE) (\ref{BAE}) is the condition of $t_1,
\cdots, t_m$ and $v = g(\lambda)e_0$, so that (\ref{eigenvector}) is an
eigenvector. The BAE is determined by the vanishing of the unwanted terms
\cite{S}. When the diagonal $L_k^k(w)$ in $T(w)$ move through the lowering
$L_j^i(t)$ to $v$, the contribution of different diagonal $L$ should be
cancelled.

The noncommutativity causes the shift of $\lambda_i$ by $\eta$. So besides
the usual B.A. condition for $t_i$ and $\lambda_i$, we following
\cite{FV2} (cf. their proof of Lemma 3 for detail), find that as the $l =
1$ case
\begin{equation}
g(\lambda) = \prod_i e^{\zeta_i \lambda_i} \prod_{j , k = 1}^n
\frac{\theta(\lambda_j - \lambda_k + \eta)}{\theta(\lambda_j -
\lambda_k)}.
\end{equation}

Check carefully for the vanishing of unwanted terms, the second factor of
$g(\lambda)$ ensures that the first $\lambda$ dependent factor of the
$V_n$ matrices $L_i^i$ and $L^j_j$ (\ref{L-intertwinner}) scattered
through $L_j^i$ be cancelled. Meanwhile the difference action on ${\cal
L}(u)$ of $L^i_i$ and $L^j_j$ shift the $e^{a\lambda}$ factor differently,
with a phase shift $e^{\zeta \eta}$, which is independent of $\lambda$.
This independence of $\eta$ is sufficient to ensure the usual B.A. eq. for
different $t_i$. But meanwhile the wave function will have the form
\cite{FV2}
\begin{equation}
\psi = \prod_i e^{\zeta_i \lambda_i} \prod_{j = 1}^n \theta(\lambda_i +
t_j - \eta),
\end{equation}
which will be twisted by $\zeta_i \eta$ when $\lambda_i$ changed by the
difference action of noncommutative Wilson loop $U_1$, $U_2$ i.e. by
diagonal $L_i^i$ and nondiagonal $L_k^j$ respectively.

In case of XYZ or equivalently the $sl(2)$ IRF, in the Gaudin limit
\cite{ST1}, the parameter $\zeta_j \equiv 2 \pi \nu_j$, $(0 \leq nu_j \leq
\frac{1}{2 \pi i \eta}$ is related to the twist in the boundary actions
\cite{S}. It describes the twist of the wave functions \cite{FT, ST}
\begin{equation}
\psi_\nu = \sum_{a \in {\bf Z}} e^{2 \pi i \nu \eta} \psi_{\nu + 2 a
\eta}.
\end{equation}
To ensure the convergence of the series in case $\eta$ equals some
rational time of $\frac{l + m \tau}{n}$, choose $\nu = \frac{Z}{n}$, then
the Fourier series truncated to finite $n$ terms \cite{FT}, and we will
obtain cyclic representation as describe in \cite{HSY}.

For the $\eta$ general case, the zero-weight condition restricted the
number $m$ of $L_{j_i}^{k_i}(t_i)$, which will be the same as zero-weight
space $nl$, (in our case $l = 1$). In case $n = 2$, this implies the
number of lattice $N$ should be $2$ $(n = 2$) times the number of
operators $L_j^k$ (their $B_j^i$). The above are arguments in usual
algebraic B.A. on vector space $V_n$ without the difference operator on
${\cal L}(u)$, which we turn now. Now the finiteness of the dimension is
the truncation of the Verma module to its quotient space \cite{FV1}.

As our discussion for QHE, we choose $\zeta_i$ to be an integer $A$, (we
have normarlized $A$ to be $1$, i.e. case $l = 1$), this will give us the
wanted noncommutativity. Actually this is just the contribution of
difference operatoin when $U_1$ like $L_i^i$, $L_{i+1}^{i+1}$ commute with
$U_2$ like $L_{i+1}^i$.

\section{Discussion}

\indent

There is a lot to be investigated later, e.g.: How to use the Weyl
Moyal transformation to give the explicit functional expressions of the
soliton solution i.e., the projection and ABS operators; how the affine
Weyl reflections imply for the duality and Morita equivalence. It is easy
to find ADE orbifold and BC orientfold case for Gaudin and C.M. models.

For the $\eta \neq 0$ case, the N.C. $q$ KZB and $q$ Virasoro should be
considered. The N.C. R.S. as C.M. can be obtained as Hitchin systems from
the cotagent bundles on N.C. torus.

Thus in anticipation for the application in N.C. models, we collecte some
known elliptic algebra in the second part of this paper.

The Sklyanin algebra is the deformed $su_n({\cal T})$. (The differencial
operator is deformed into difference of $\eta$ ). The Lax pair and
$R$-matrix of $Z_n \otimes Z_n$ Belavin is the deformed $L$-matrix and
$r$-matrix of elliptic Gaudin. The cotangent bundle on torus for algebra
will be deformed into that for group \cite{ACF}. The C.M will be deformed
into RS related with IRF and the quantum group.

The corresponding field are the gauge WZW and CS \cite{GN}, the N.C.
case has been studied recently by \cite{GKK}.

We will investigate more carefully N.C. aspect of the integrable many-body
and gauge field \cite{GM} and string brane \cite{M} later. The dynamical
variables $z_i (\sim \lambda_i)$ and $\partial_i$ in exponents will be the
$z_i$ and $\bar{z}_i (\sim \partial_i)$ for N.C. soliton. The N.C. source
now equals the $U_1 U_2 U_1^{-1} U_2^{-1}$ as the affine Heisenberg Double
\cite{AFM} and the N.C. Wilson loop for the RS reduced form CS \cite{GKK}.

\end{document}